**TITLE:** Rapid Prototyping Model for Healthcare Alternative Payment Models: Replicating the Federally Qualified Health Center Advanced Primary Care Practice Demonstration[1]

**Running Head:** Rapid Prototyping Model for Healthcare APMs

**Authors:**
- Jarrod Olson, MPP[2] (corresponding author) (Battelle Memorial Institute, Seattle, WA 98109, olsonjr@battelle.org)
- Amir Rahimi, PhD[3] (Battelle Memorial Institute, Columbus, OH 43201, rahimia@battelle.org)
- Po Hsu Allen Chen, PhD[2] (Battelle Memorial Institute, Columbus, OH 43201, chenp@battelle.org)
- J. Elizabeth Jackson, PhD[1] (Battelle Memorial Institute, Seattle, WA 98109, jacksonje@battelle.org)
- Tyler Coy, BS[2] (Battelle Memorial Institute, Columbus, OH 43201, coy@battelle.org)
- Adrienne Cocci, MPH[3] (Battelle Memorial Institute, Arlington, VA 22202, cocci@battelle.org)
- Nancy McMillan, PhD[2] (Battelle Memorial Institute, Columbus, OH 43201, mcmillann@battelle.org)
- Jeff Geppert, J.D., Ed.M.[2] (Battelle Memorial Institute, Columbus, OH 43201, geppertj@battelle.org)

[1] **Financial Support Statement:** Financial support for this study was provided entirely by a grant from Battelle Memorial Institute's Independent Research and Development program. The funding agreement ensured the authors' independence in designing the study, interpreting the data, writing, and publishing the report. All of the authors are employed by the sponsor.
[2] Battelle Memorial Institute, Seattle, WA 98109
[3] Battelle Memorial Institute, Columbus, OH 43201
3 Battelle Memorial Institute, Arlington, VA 22202

# Abstract

**Background and Objective.** Innovation in healthcare payment and service delivery utilizes high cost, high risk pilots paired with traditional program evaluations. Decision-makers are unable to reliably forecast the impacts of pilot interventions in this complex system, complicating the feasibility assessment of proposed healthcare models. We developed and validated a Discrete Event Simulation (DES) model of primary care for patients with Diabetes to allow rapid prototyping and assessment of models before pilot implementation. We replicated four outcomes from the Centers for Medicare and Medicaid Services Federally Qualified Health Center Advanced Primary Care Practice pilot.

**Methods**. The DES model simulates a synthetic population's healthcare experience, including symptom onset, appointment scheduling, screening, and treatment, as well as the impact of physician training. A network of detailed event modules was developed from peer-reviewed literature. Synthetic patients' attributes modify the probability distributions for event outputs and direct them through an episode of care; attributes are in turn modified by patients' experiences.

**Results.** Our model successfully replicates the direction of the effect of physician training on the selected outcomes, and the strength of the effect increases with the number of trainings. The simulated effect strength replicates the pilot results for eye exams and nephropathy screening, but over-estimates results for hemoglobin A1c (HbA1c) and low-density lipoprotein (LDL) screening.

**Conclusions.** Our model will improve decision-makers' abilities to assess the feasibility of pilot success, with reproducible, literature-based systems models. Our model identifies intervention and healthcare system components to which outcomes are sensitive, so these aspects can be monitored and controlled during pilot implementation. More work is needed to improve replication of HbA1c and LDL screening, and to elaborate sub-models related to intervention components.

# Introduction

Since Medicare and Medicaid (now administered by the Centers for Medicare and Medicaid - CMS) were established in the 1960s, reimbursement for health services has used the Fee-For-Service (FFS) payment model, where CMS reimburses providers based on inputs to the care process (e.g. number of clinician hours per procedure) "rather than the most appropriate care over an episode of illness or over the course of a year".[1(chap14 pg 3)] FFS incentivizes healthcare providers to perform a large volume of services, even when better, lower-cost methods exist to treat a condition, leading to higher cost, lower quality healthcare. With bi-partisan support, payment and service delivery reform using Alternative Payment Models (APMs) started in 1988 when Medicare was authorized by Congress to test a new reimbursement model, the Coronary Artery Bypass Graft (CABG) bundled payment. Another model was tested in 2005 looking at Physician Group Practice payment models. Both payment reforms tied reimbursement to performance against clinical quality measures (CQMs) to ensure that lower costs were not achieved at the expense of quality of care.[2] In the 2009 Affordable Care Act, Section 3021 created the Center for Medicare and Medicaid Innovation (CMMI) and gave it broad authorization to test new models that tie payment to performance against CQMs or implement changes to service delivery models. CMMI is authorized to test models brought forward by legislation, the Physician-Focused Payment Model Technical Advisory Committee (PTAC), or other sources based on the judgment of the Secretary of Health and Human Services (HHS)[3]. In the 9 years since its creation, CMMI has planned or executed more than 70 new APM[4] models aiming to either decrease cost, increase quality, or both.

In that time, only 12 models have been completed through the evaluation phase. As of 2016, only 2 were certified for national adoption by Medicare's Chief Actuary, based on the criteria that the pilot reduced cost, increased quality, or did both.[4] Each of those 12 pilots took a minimum of 3 years and exposed the government (as the funder), providers (as the implementers), and patients to risk by changing the way that healthcare is provided at a system level. The direct cost of running pilots is high, with CBO estimating that CMMI will spend $12 billion from 2017-2026, and the time to evaluate them can stretch into multiple years.[4(p3)] As an example, the Federally Qualified Health Center (FQHC) Advanced Primary Care Practice (APCP) Demonstration final

---

[4] APM is used throughout this paper to reflect both payment and service delivery models.

evaluation report was submitted to CMS 5 years after demonstration start (started in 2011, submitted in September 2016) and published in 2018.[5] Our analysis of evaluation methods used in the 12 completed APM evaluations (as of Spring 2017) showed that evaluations typically leveraged a Difference-in-Differences (DiD) approach using propensity score matching to estimate the impact of the demonstration on outcomes, and utilized randomized treatment and control groups less often (see Woolridge[6(p147)] for discussion of DiD). Analysis of the implementation and resulting organizational changes were almost exclusively qualitative and based on interviews and focus groups. Many of the evaluations showed limited impact quantitatively and reported cost savings that were primarily driven by re-negotiated reimbursement rates.

The time, cost, and risks of the current framework for APM evaluation are too high and have not produced substantive improvements for CMS. Rapid prototyping of APMs using a simulation model before patients, providers, and the government are put at risk could minimize those costs and risks and increase CMMI's "win-rate" by filtering out unsuccessful concepts and providing useful information to implementors and evaluators. This paper introduces the Rapid Prototyping Model (RPM), a Discrete Event Simulation (DES) designed to help CMMI meet their pilot decision-making needs. RPM implements the healthcare system as a series of events that a simulated population of patients, clinicians, and organizations all experience based on a set of probability distribution functions (PDFs) characterizing a set of outputs for each event. The use of a probabilistic simulation model at the patient level means that we can test a broad range of assumptions by configuring the patient population, the structure of the model, and runtime considerations. We use RPM to replicate CMMI's FQHC APCP demonstration, calculating the demonstration's quantitative measures for testing. We show how the results can provide critical information about the areas of greatest implementation uncertainty, including how implementation scenarios can impact results.

# Methods

### Selection of Simulation Method

Three primary microsimulation modeling paradigms have been defined for modeling complex systems composed of entities of various types - firms (e.g., factories, warehouses) or human actors (e.g., retailers, customers) - where the desire is to capture dynamic, macro-level change arising from the interactions of micro-level entities.[7] These three paradigms– Systems Dynamics (SD), Discrete Event Simulation (DES), and Agent-Based Modeling (ABM)– differ primarily in the heterogeneity and detail of the modeled entities' attributes, the fixedness of the system or process modeled, and the degree to which entities are passive or active participants in the system. Each feature impacts the types of hypotheses that can be investigated. These paradigms were developed primarily to address problems in management and engineering,[8] but each has been employed to varying degrees to model the likely impacts of changes in health policy [9–15]. Systems models like the three noted here are especially useful for the study of unintended consequences, where an intervention may result in a side-effect that negatively impacts population health. For example, increased access to care may lead to increased utilization of care, and a shortage of clinicians, serving to make it more difficult for patients to access care when needed[16].

DES is a paradigm that focuses on replicating real-world processes and depends on a detailed flow of events where resources possessing attributes pass through the flow based on event outcomes that are influenced by the resource's attributes. It has been applied broadly outside the

U.S. for healthcare (e.g. Lebcir et al.[9]). Although SD and ABM have been used to address questions in the healthcare area, SD tends to overly aggregate the outcomes of interest and ABM requires extensive bottom-up development of agents, incurring high costs to justify "rapid prototyping".[7]

DES, with its top-down approach to system complexity and the ability to model heterogeneous populations provides an excellent framework for the RPM. In addition, experimentation in DES is conducted by changing the process structure, much like interventions are intended to change healthcare processes. Finally, its emphasis on queuing processes and resource constraints can be leveraged to address common issues in access to care that are the focus of demonstrations such as the FQHC APCP. In the current study, we apply the DES approach to healthcare APMs to assess its use for pre-pilot feasibility assessments.

**The RPM Framework**

RPM is a Python-based DES leveraging the SimPy[17] library as an asynchronous event dispatcher and pseudo-random number generation using the NumPy 'random' function.[18] It explicitly models resource constraints (including insurance) and queuing to explore the dynamic nature of healthcare, meaning that interventions addressing access to care and other resource limitations are also amenable to evaluation using this model. This resource constrained approach differentiates RPM from the similar Archimedes model, also used to study CMS policy impacts on Diabetes treatment,[19,20] which assumes that resources are unlimited and immediately available, and does not account for healthcare insurance in its population[21(p56)]. Similar models using SD often model congestion using a supply and demand mechanism, but cannot provide detail about the impacts by patient or clinician.[10,11]

Following DES conventions, RPM is composed of a population generator and a simulator. The population generator utilizes the same logic and probability distribution functions (PDFs) as the simulator to generate a patient pool with attributes drawn from pre-specified distributions. These attributes modify the probability distribution of event outcomes in the simulator. **Because the events are probabilistic, we can create placeholder events with a uniform probability distribution when we have no prior assumptions about its PDF with a cost of increased uncertainty in the final expected value**. We can add complexity to the model iteratively by adding greater detail to event PDFs. The development process consists of three steps: 1) specification (identifying the literature, assumptions, and algorithms to define an event); 2) parameterization (assigning values to model specification); 3) and implementation (integrating specification and parameterization into the code base).

Specification is the process of identifying events using a narrative, descriptive perspective, identifying the attributes for the population, and identifying the algorithm underlying the PDF. This process is the most time-consuming part of model development. A team of health services researchers defined a set of events and pathways between them, and then built the RPM Event Library and Attribute Library one event at a time by identifying and documenting literature about the event and corresponding attributes, algorithms for the outcome probability distribution, and assumptions driving the algorithm.

Parameterization is the process of converting narrative event and attribute descriptions into PDFs. A team of statisticians using R[22] worked with the RPM Attribute Library to identify the distribution of the attributes in the population and incorporate that information into the model. Parameterization also includes operationalization of the event specifications and fitting probability distributions to the summary statistics reported in journal articles. In a few cases, it required the team to estimate a probability distribution from existing data and to correlate these distributions when there were dependencies.

Implementation is the process of turning the parameterized event and attribute specifications into an actual Python simulation using the SimPy library. This task was led by a team of computer scientists. The model was designed in such a way that events are modular and can be added or removed based on the scenario in question.

Uncertainty characterization is an integrated element of the model development process. The team specified probability distributions, rather than point estimates for inputs to the model to support Monte Carlo analysis. Sets of these distributions define a scenario quantitatively (i.e. a set of assumptions) and can be compared for differences.

For Quality Assurance, we utilized three processes. First, automated integration and unit testing of the software ensured high quality code. Second, the team manually reviewed simple, but realistic scenarios using pre-specified hypotheses to assess event-scenario outputs and confirm that general behaviors worked as expected. Third, corner-case validation used specific scenarios that isolated components of the model. Each of the three processes was documented and scored to be provided as a simulator artifact with the version-controlled (using Git) release of the software.

## Selection of Model to Replicate and RPM Specification

We selected a completed CMMI APM to validate our modeling approach using three primary criteria: 1) it had significant findings of intervention effect so that there was a non-zero impact to replicate, 2) there was a sufficiently detailed description of the expected implementation from which we could develop an event flow, and 3) the findings were reported with uncertainty so that we could assess overlap with our model outputs.[23] We reviewed and abstracted data for 12 completed models (as of Spring 2017) in detail to assess their viability for replication and were able to identify <u>only one</u> with a reasonable chance of replication.

The model we selected for replication, the FQHC APCP, met the criteria described above: (1) four significant effects on CQMs related to Diabetes, (2) a clear conceptual model and a set of defined intermediate objectives that were theorized to lead to improved outcomes, and (3) the results were reported with standard errors.

The pilot intervention helped participants achieve the Patient-Centered Medical Home (PCMH) Level 3 Designation. PCMH Level 3 requirements for an APCP correspond to the intervention elements in Table 1 in column 1. The model implementation in Table 1 column 2 is discussed in detail below. Once PCMH Level 3 designation is achieved, the structural changes reflected by this designation are expected to improve clinical quality and patient outcomes.

[Table 1 about here]

The event flow for this study consists of 24 core events requiring 53 attribute parameters summarized into a high-level diagram in Figure 1 (shown in detail in Appendix 1). The RPM FQHC APCP model simulates patients' daily lives with a prompt for entry into the simulation ("due for a wellness check," "getting sick," and "developing diabetes symptoms"). These critical pathways form the boundary of the static, one-year simulation for patients and each episode of care includes scheduling and the doctor visit. The schedule is a first-in, first-out queue following the "Simplified Scheduling Process."[24] The intervention is modeled as organizational change where doctors receive training and alternative scheduling approaches are utilized, leading to a change in care delivery. The specifics of the intervention are discussed in Table 1, column 2 "Model Implementation." Initial conditions are simulated during warm-up to pre-populate regularly recurring appointments and the population generator specifies initial health statuses.

[Figure 1 about here]

**Replication Design and Simulation**

The overall approach for replication utilizes literature and data available in 2011, **before pilot implementation, to reproduce the likely real-world conditions of simulating this pilot prior to its implementation**. The literature and data were used to parameterize and build necessary sub-models aligned with the changing features in primary care from the pilot.

For the replication experiment, we generated two scenarios. The first was the "baseline" case with a standard FFS FQHC system. The second, "pilot" scenario assumed that the demonstration was implemented "as expected," in terms of Table 1. Table 1 shows the expected change for an APCP-designated facility and the corresponding element of the intervention scenario from RPM. The simulator implements only changes that are expected to substantively change the results. For element 1, we specify a sensitivity case of 1-5 trainings to understand how the intervention implementation may impact results. We assumed that 5 one-week trainings were a reasonable upper bound that would likely exceed a reasonable real-world upper bound for clinician trainings in a year. Due to design constraints, intervention elements 3-5 will be tested in future work, and element 2 was partially implemented, but those changes are not expected to substantively change the result.

The simulation study compares the outcomes from the baseline model with the outcomes from the pilot model to determine the expected impact on the CQMs that the pilot aims to impact (shown in Table 2). Although the pilot utilized more than 50 clinical quality measures, we focused on 4 claims-based measures reflecting guideline-recommended screenings for patients with Diabetes that were sensitive to the intervention. The simulated impacts are compared against corresponding pilot study impacts as reported in the final evaluation report.[23]

[Table 2 about here]

Because it is computationally expensive to model all the 1,198 FQHCs, we leverage similarities in FQHC-attributes to empirically cluster FQHCs into common "types" based on patient attributes such as mean age, socioeconomic status, racial/ethnic diversity, insurance access, and overall population size served using the Health Center Program Grantee Profiles dataset[5] (data reflecting 2011 state of grantees) (see Table 3). We then run our pre- and post-intervention scenarios for each of the four clusters separately and combine the results to generate a national estimate of the expected intervention impact using a weighted ANOVA model with weights assigned by the number of FQHCs in each cluster. Each scenario is tested for a one-year period, with 200 runs per cluster per scenario (5 intervention scenarios to account for 1-5 doctor trainings).

[Table 3 about here]

We compared our simulated impacts against actual impacts reported in the pilot evaluation using a comparison of the confidence intervals for the real-world data and for the simulated data. We compare the simulated change against each of the three years of actual results. A CQM result is considered successfully validated if its actual 95% confidence interval overlaps with the simulated 95% confidence interval.

---

[5] Accessed May 2017 at https://bphc.hrsa.gov/uds/datacenter.aspx?q=d&year=2011

This study was funded with independent research and development funds from the authors' institution.

# Results

Comparing the overall RPM impacts with the actual impacts, it appears that the simulator (averaged across all simulator runs) over-estimates the effect of the intervention. Figure 1 shows the comparison of the RPM-modeled intervention effect to the 3-years of reported intervention effect from the actual pilot. The accuracy of the simulated results is represented by overlapping confidence intervals between the RPM simulated effect and the pilot actual effect. In the plot, eye exams and nephropathy are closest to replication, but the confidence intervals do not consistently overlap.

[Figure 2 about here]

Results controlling for the number of trainings are shown in Figure 2. The importance of the details of model implementation are clearly illustrated, where the number of trainings provided to clinicians changes the simulated intervention effect significantly. Training effect is drawn from a systematic review that shows the impact of one additional period of substantial continuing education (e.g. attending a training ranging in length from 1-5 days) on a clinician's overall performance related to CQMs.[25] The RPM estimates demonstrate that the effects of different training levels is both significant and critical to the performance of the APM (See Figure 2). The estimates for Eye Exams and Nephropathy with one or two trainings per staff indicate a successful replication with overlapping confidence intervals. On the other hand, the estimates for LDL and HbA1c were too optimistic.

[Figure 3 about here]

In addition to the finding that number of trainings is a significant differentiator in the intervention, cluster factors were significant differentiators in the RPM model results. Three of four clusters were significant in the models for HbA1c, LDL, and Eye Exam CQMs, while cluster factors were not significant in the model for Nephropathy. Cluster factors essentially capture demographic and socioeconomic factors. Availability of care was not significantly impacted in any of the models.

## Discussion

Our model successfully replicates the direction of effect of physician training on the selected outcomes, and the strength of effect increases with the number of trainings. The results show that APM implementation details are critical to understanding expected impact. Specifically, the number of trainings provided to clinicians significantly impacted the simulated success of the intervention. The demonstration of model sensitivity to parameters demonstrates the value of using rapid prototyping to assess the proposed APM implementation by identifying training as a key sensitivity. Our results showed that a requirement that organizations provide 2 or more in-depth trainings might increase the success of the intervention. In designing an evaluation, data collection about the number of clinician trainings, data not reported in the real-world evaluation, could help to validate and refine this finding for future efforts.

The RPM was developed exclusively from existing studies and datasets, requiring no first-hand data collection and exposed no patients, providers, hospitals, or payers to risk. This model showed that the proposed intervention would likely yield benefits for all of the four CQM tested here and demonstrated that the impact of the intervention on each is sensitive to how it is implemented. As noted above, this information is useful to the intervention designers, as they can more concretely specify guidance and expected payoff for the way the intervention is implemented. Additional work is needed to explain the differences in the intervention effect for HbA1c and LDL. Both the original pilot study and the simulated results show a positive, generally significant effect. Replicating magnitude from the simulation model was less reliable. Additional data regarding the amount of supplementary training given by Level-3 PCMHs would be needed to validate the successful eye exams ordered and nephropathy testing results.

The results of the RPM study are useful in traditional evaluation approaches because the simulated results help prioritize areas where quantitative data is most needed by identifying gaps in the scientific record, or areas of implementation that could lead to major differences in outcomes. In the real-world study this is based on, the evaluators could only use pre-study and post-study outcomes and explained insignificant or counter-intuitive results with qualitative analysis. If grantees reported more granular implementation information, like the number of clinician trainings provided, it might be enough to explain some variation in the results across pilot sites. Another source of variation is the type of FQHC, which is largely determined by a set of demographic and socioeconomic characteristics of patients. With this information, the implementers and evaluators could more concretely specify comparison groups or design matching algorithms to control for that variation. CMMI model participation is competitive and participants opt-in to the pilot. The use of a DiD approach using propensity score matching builds a more directly comparable control group, but if those matching characteristics could be identified *a-priori* from empirical simulated data, it would provide greater confidence in the matching.

RPM demonstrates the value of DES for rapid prototyping but does have some weaknesses to address in future development. First, the model was static for this analysis, relying on a very rigidly specified set of pre- and post-intervention assumptions. This fails to account for clinician and organization implementation path and may be driving the higher impact than the reported real world. Nor does it capture the life-course of the patient and patient population over time. This likely overstates the model results by reducing variation. The RPM model is slated for a dynamic patient, clinician, and organizational development model in the next phase of development. This will make it possible to account for unintended consequences in the system and will offer a richer set of results and conclusions.

Another issue is that we assumed a linear effect for the impact of trainings that is generalized as "substantial training" from a systematic review. It may not be the most appropriate representation of the intervention setting and targeted outcomes. In future research, alternative models for the impact of training should be considered.

RPM was developed with a general population derived from descriptive statistics to specify the population characteristics. Ideally, we would have used real-world patient data. This would more richly capture the interactions between patient chronic conditions, socio-demographic information, and other elements that would help reflect the real-world settings and result in the findings about unintended consequences.

One of the biggest challenges to building RPM was finding the literature to support event specification. As an example, the "HbA1c Screening Test Ordered" module currently leverages a 2005 study in Oregon about a single FQHC's HbA1c screening rates[27] because other, more general population studies were not available. That paper had lower baseline rates than the FQHC APCP evaluation report. Many quantitative studies were unusable in their published form because they did not report all variables ("the controls") and rendered their quantitative results useless for simulation. One approach to resolve this issue is original research, and we hope that this model can drive new areas of investigation. In some cases, the screening and diagnostic processes foundational to primary care do not appear to have been published. The workflow and processes for treatment are often very narrowly focused on a single drug or lifestyle intervention, making it difficult to incorporate this information into a model.

## Conclusion

This paper demonstrates that the RPM DES model can be used to assess the feasibility of new APM models and to inform the design of those models for evaluation and implementation. This initial model was built using only available literature and datasets, with minimal secondary data analysis. Using this foundation, our model replicated 2 of the 4 measures from a real-world pilot, assuming an implementation of only one or two trainings per clinician per year to improve their performance. Our findings provide a stepping stone for further improvement of future model development. Additional work is needed to better understand the underlying probabilities of being tested for HbA1c and LDL compared to the simulated results.

Our findings may inform CMMI and other healthcare policy makers and organizations with APM reform to better design and evaluate the pilots that they implement to improve quality and reduce the cost of healthcare. The current approach of developing a new intervention over several years, recruiting participants using an opt-in, (usually) non-randomized approach, then placing patients, providers, and payers at risk during a three-year pilot, while waiting multiple years for evaluation results, does not allow the system to "fail fast" and incurs substantial costs. Additionally, without prioritized data collection that measures variation in implementation across key intervention elements, it is difficult for evaluators to estimate anything more nuanced than an overall pre- and post-intervention effect. Simulation tools allowing for rapid prototyping, such as DES, are a promising innovation in the design and evaluation of APMs and could yield substantial process improvements and cost savings by avoiding unproductive pilots, and prioritizing data collection for evaluators.

# Acknowledgments

We would like to acknowledge the research support from our team, including Amira Elhagmusa, April Greek, Hyoshin Kim, Mary Sheehan, Cheryl Triplett, and Jamie Turner. We would also like to acknowledge the members of our advisory committee, including Wanda Gamble, Liz Herman, Kara Morgan, and Warren Strauss.

# Tables

Table 1 – Intervention Elements and Scenario Implementation for RPM FQHC APCP model

| Intervention Element (From APCP Specification) | Model Implementation |
|---|---|
| General improvement through systems change and provider training | Clinicians attend randomly assigned number of trainings per year and receive a corresponding increase in performance based on a systematic review showing a mean effect of 14 percentage points with a 95% confidence interval from .68 to 26.44 percentage points. We implement this as a proportional change from the baseline with a random draw across the 95% confidence interval range [25] |
| FQHC hours are changed to increase availability (leading to longer hours and staggered shifts) | We incorporate longer hours and stagger shifts in a sensitivity case but see no significant differences for the CQMs in this replication, probably because patients do not prefer a time of day for appointments in the simulation. For brevity, these results are not reported. |
| More same day appointments are available | Not implemented, see note about no utility function for patient appointments. If an appointment exists within a time range, patients will accept. |
| Assigning patients to individual clinicians | Implemented in both baseline and intervention. |
| Creation of registries to improve communication about patient issues | Implemented in both baseline and intervention. |

Table 2 - Clinical Quality Measures (CQM) Difference-in-Differences Results to replicate from FQHC APCP

| CQM | Numerator | Denominator | Difference in Differences (%) | | |
|---|---|---|---|---|---|
| | | | Year 1 | Year 2 | Year 3 |
| **HbA1c testing for diabetic patients** | Patients with HbA1c Tests Performed | Patients with diabetes | 1.67 (.43) | .68 (.102) | .70 (.38) |
| **Eye examination for diabetic patients** | Patients with Eye Examinations performed | Patients with diabetes | 1.84 (.50) | 1.17 (.47) | 1.23 (.46) |
| **Nephropathy testing for diabetic patients from claims** | Patients with Nephropathy Tests performed | Patients with diabetes | 2.62 (.55) | 3.36 (.51) | 2.62 (.49) |
| **LDL testing for diabetic patients from claims** | Patients with LDL Tests performed | Patients with diabetes | .48 (.330) | .16 (.728) | 1.00 (.46) |
| Values reported as percentage point differences and standard errors from the evaluation report[23(p204)] | | | | | |

Table 3 – Cluster Attributes and Number of FQHCs

| Cluster | Attributes | Number FQHCs |
|---|---|---|
| 1 | Older, low diversity population, with better access to insurance, and more income | 399 |
| 2 | High diversity, poor, with a lot of uninsured, in FQHCs serving a small population | 274 |
| 3 | Poor, with higher diabetes, relatively high diversity (~50% minority), and FQHCs serving a large population | 69 |
| 4 | Young, relatively poor, with high diversity, and high Medicaid enrollment and FQHCs serving a large population | 456 |
| Clusters generated using GAP software[26] from BPHC Health Center Program Grantee Profiles in 2012 | | |

# Figures

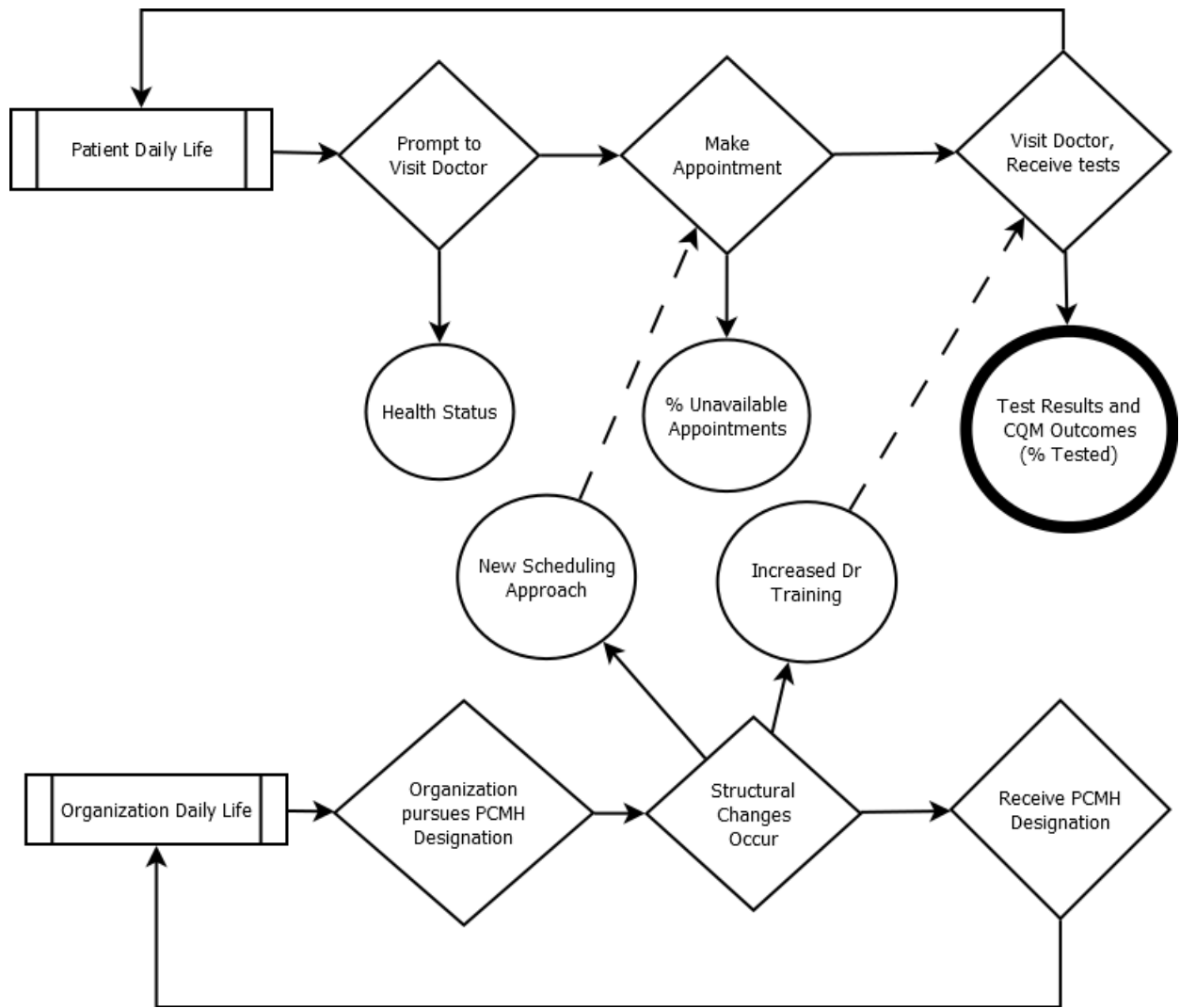

*Figure 1 – High level flow depicting the RPM FQHC APCP Model Event Flow (diamonds are events, circles our outcomes, replication outputs in bold circle)*

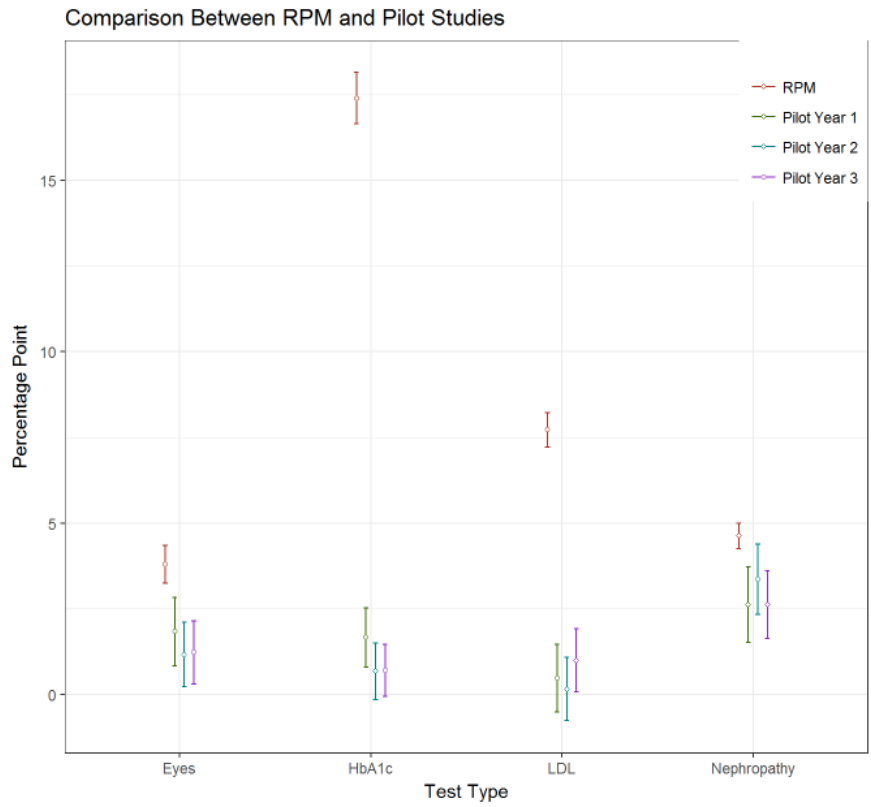

*Figure 2 – Comparison of RPM Results and Pilot Results (95% Confidence Intervals reported for 4 test types)*

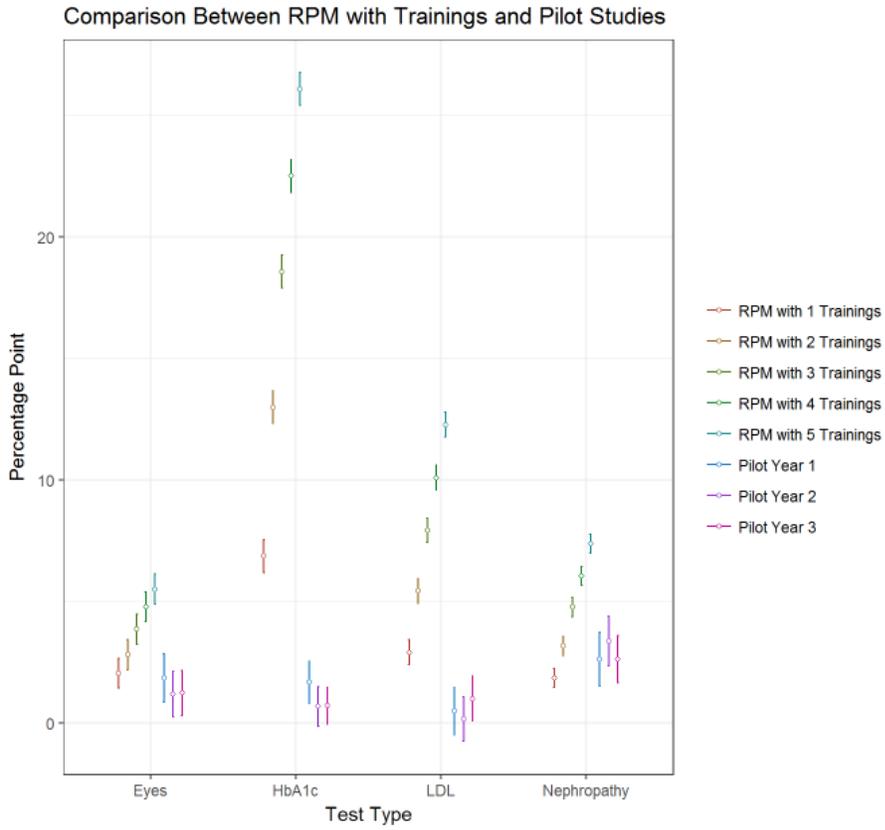

*Figure 3 – Comparison of RPM and Pilot Results, faceted by number of trainings (95% Confidence Intervals reported for 4 test types)*

# Appendices

## Appendix I – Detailed Flow Diagram for RPM

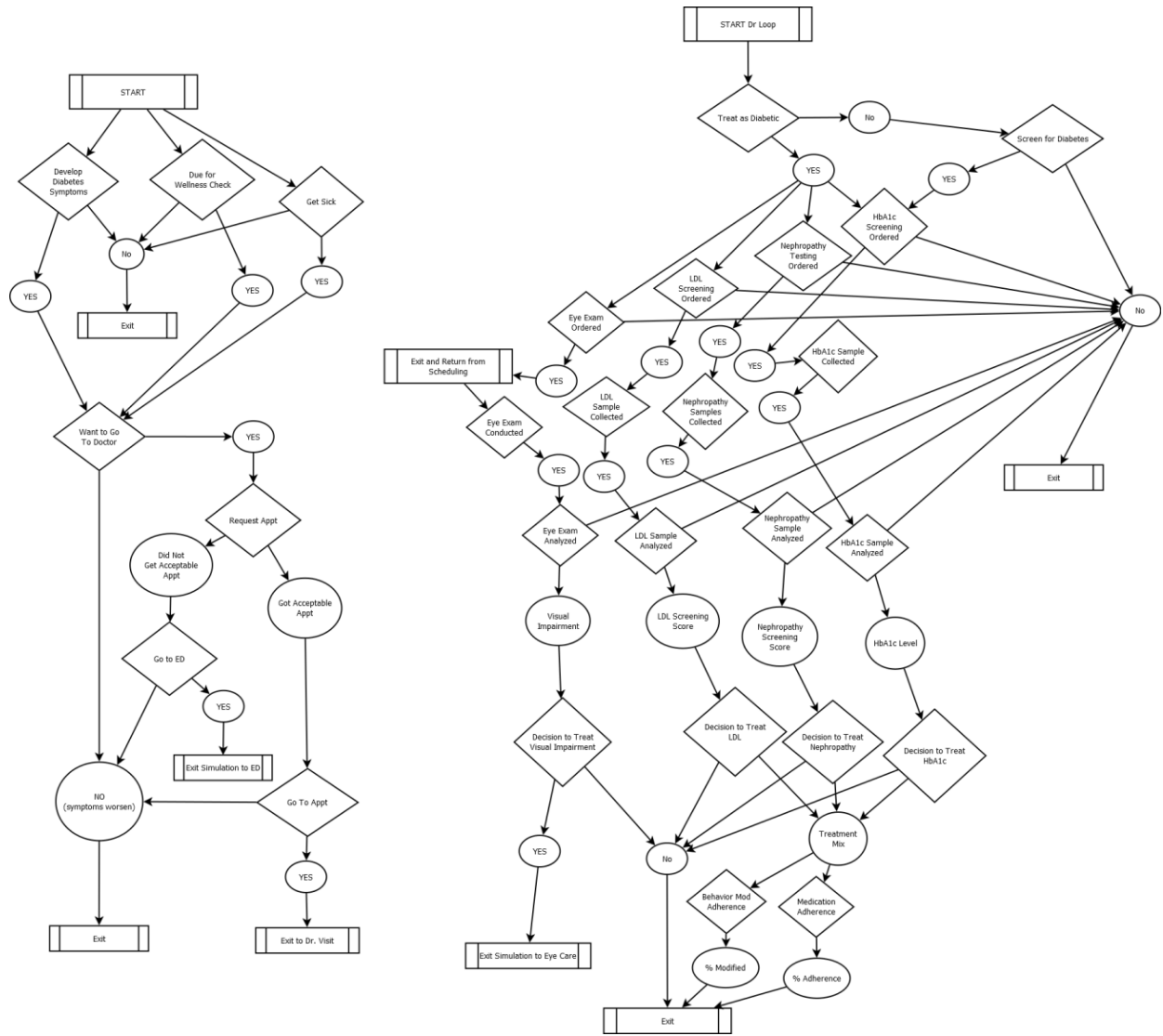

*Figure 4 – Detailed flow-diagram of all events and outputs for the RPM simulator. Diamonds represent events, and circles are outcomes. For simplicity, we break the visualization into two core modules "Scheduling" and "Dr. Visit".*